\documentstyle[proceedings,numreferences]{crckapb}

\def\openone{\leavevmode\hbox{\small1\kern-3.8pt\normalsize1}}%

\def\overcirc#1{\protect
\setbox0=\hbox{$\displaystyle #1$}%
\setbox1=\hbox{$\scriptstyle \circ$}%
\setbox2=\hbox{}\ht2=\ht0 \dp2=\dp0 %
\ifdim\wd0>\wd1 %
\setbox1=\hbox to\wd0{\hss\box1\hss}%
\mathord{\rlap{\raise1.2\ht0\box1}\box0}%
\else   %
\setbox1=\hbox to.9\wd1{\hss\box1\hss}%
\setbox0=\hbox to\wd1{\hss$\displaystyle\relax#1$\hss}%
\mathord{\rlap{\copy0}\raise1.2\ht0\box1}%
\fi
\mathord{\box2}}

\begin{opening}

\title{Chiral Interactions of Massive Particles
in the $(1/2,0)\oplus (0,1/2)$ Representation}
\author{Valeri V. Dvoeglazov}

\institute{ Escuela de F\'{\i}sica, Universidad Aut\'onoma de Zacatecas \\
Apartado Postal C-580, Zacatecas 98068, Zac., M\'exico\\
Internet address:  valeri@cantera.reduaz.mx\\
URL: http://cantera.reduaz.mx/\~~valeri/valeri.htm}

\runningtitle{CHIRAL INTERACTIONS}
\runningauthor{Dvoeglazov}

\end{opening}

\begin{document}

%\begin{center}
%Submitted 25 August  1997
%\end{center}

\begin{abstract}
On the basis of the first principles we argue that self/anti-self
charge conjugate states of the $(1/2,0)\oplus (0,1/2)$ representation can
possess the axial charge. Finally, we briefly discuss recent claims
of the $\sim \vec \sigma \cdot [ \vec A \times \vec A^\ast ]$ interaction
term for the particles of this representation.


\end{abstract}

%\smallskip

It is well known that the Dirac equation and the relevant theory of
charged particles do {\it not} admit the $\gamma^5$ chiral transformations.
The sign in the mass term in the Lagrangian is reversed under this type of
transformations. In the mean time, the chiral transformations play
significant role in our understandings of the nature of weak and strong
interactions, in the problem of (un)existence of monopoles as well.
In the present article we bother with the McLennan-Case
reformulation~\cite{MLC2,MLC1} of the Majorana
theory~\cite{MAJOR}, investigate the possibility of
the different choice of the field operator and
prove that massive self/anti-self charge conjugate states appropriately
defined in the $(1/2,0)\oplus (0,1/2)$ representation possess the axial
charge (cf. also refs.~\cite{DVA,ZIINO1,ZIINO2,DVO,ZIINO3} and
related papers~\cite{DVA0,DVO0}).

We start from the observation that the Dirac field operator,
which satisfies the Dirac equation\footnote{Of course, the
mathematical framework, which we present, depends on the
definitions and postulates. One can start from the Dirac equation
or one can start from the postulates of D. V. Ahluwalia~\cite{BWW},
see below. But, the physically relevant conclusions do not change. The
notation and metric of the paper~\cite{MLC1} is used.
Namely, $g^{\mu\nu} = \mbox{diag} (-1, 1, 1, 1)$ and
$\gamma$ matrices can be chosen as follows $$\gamma^0 =\pmatrix{0&-i\cr
-i&0\cr}\, , \, \gamma^i = \pmatrix{0&i\sigma^i\cr -i\sigma^i &0\cr}\, ,
\, \gamma^5 = \pmatrix{1&0\cr 0& -1\cr} \, .$$ The Pauli
charge-conjugation $4\times 4$ matrix is then $$C = e^{i\vartheta_c}
\pmatrix{0&\Theta\cr -\Theta &0\cr}\, , \, \mbox{where}\,\,\, \Theta =
\pmatrix{0&-1\cr 1&0\cr}\, .$$ It has the properties $$C=C^{^T}\, ,\,
C^\ast = C^{-1}\, ,$$ $$C^{-1}\gamma^\mu C = \gamma^{\mu^{\,\ast}}\, ,\,
C^{-1}\gamma^5 C = -\gamma^{5^{\,\ast}}\, ,$$ which are independent
on the representation for $\gamma$ matrices.}
\begin{equation}
[ \gamma^\mu \partial_\mu +\kappa ] \psi (x^\mu) = 0\, , \label{1}
\end{equation}
can be expanded in the following parts:
\begin{eqnarray}
\psi (x^\mu)
&=& \psi_\uparrow (x^\mu) + \psi_\downarrow (x^\mu) \,\, ,\\ \psi_\uparrow
(x^\mu) &=& \int \frac{d^3 {\vec p}}{(2\pi)^3 2E_p} \left [ u_\uparrow
(p^\mu) a_\uparrow (p^\mu) e^{-i\phi} + {\cal C} u_\uparrow^\ast  (p^\mu)
b_\downarrow^\dagger (p^\mu) e^{+i\phi} \right ] \,\, ,\\ \psi_\downarrow
(x^\mu) &=& \int \frac{d^3 {\vec p}}{(2\pi)^3 2E_p} \left [ u_\downarrow
(p^\mu) a_\downarrow (p^\mu) e^{-i\phi} - {\cal C} u_\downarrow^\ast
(p^\mu) b_\uparrow^\dagger (p^\mu) e^{+i\phi} \right ]\, ,
\end{eqnarray}
where $\phi = (Et -{\vec x}\cdot {\vec p})/\hbar$.
The charge-conjugate counterparts of these ``field operators"
are found in a straightforward way.
Both $\psi_{\uparrow}$,\,
$\psi^c_{\uparrow}$  and $\psi_{\downarrow}$,\, $\psi^c_{\downarrow}$ can
be used to form self/anti-self charge conjugate ``field operators" in the
coordinate representation after regarding corresponding superpositions.
For instance, $\Psi^S = (\psi_\uparrow + \psi^c_\uparrow)/2$ ,\,\,
$\Psi^A = (\psi_\downarrow - \psi^c_\downarrow )/ 2$ , \,\,
$\widetilde \Psi^S = (\psi_\downarrow + \psi^c_\downarrow)/2$ ,
and $\widetilde \Psi^A = (\psi_\uparrow - \psi^c_\uparrow )/2$.
As opposed to K. M. Case we
introduce the interaction with the 4-vector potential in the beginning and
substitute $\partial_\mu \rightarrow \nabla_\mu = \partial_\mu -i e A_\mu$
in the equation (\ref{1}). For the sake of generality we assume that the
4-vector potential is a {\it complex} field $A_\mu = C_\mu +i B_\mu$, what
is the extension of this concept comparing with the usual quantum-field
consideration.\footnote{In the classical (quantum) field theory the
4-vector potential in the coordinate representation is a {\it real}
function(al).  We still note that different choices of a) relations
between the left- and right- parts of the momentum-space bispinors; b)
relations between creation and annihilation operators in the
field operator; and  c) metrics would induce ones to change this conclusion
for interactions of various field configurations which one considers.}
The charge-conjugated equation to (\ref{1}) reads
\begin{equation}
(\gamma^\mu \nabla_\mu^\ast +\kappa) C \psi^\dagger (x^\mu) =0\, .
\end{equation}
Following the logic of refs.~\cite{MLC2,MLC1,ZIINO1,ZIINO2,ZIINO3}
(the separation of different {\it chirality} sub-spaces) we also consider
additional equations for $\gamma^5 \psi_{\uparrow\downarrow}$ and
$\gamma^5 \psi_{\uparrow\downarrow}^c$. Next, let us introduce the
following linear combinations
$\psi_1 = \psi_\uparrow^c - \gamma^5 \psi_\downarrow$\, ,\, $\psi_2 =
\psi_\downarrow + \gamma^5 \psi_\uparrow^c$ \, ,\, $\psi_3 =
\psi_\downarrow^c + \gamma^5 \psi_\uparrow$\, ,\, and $\psi_4 =
\psi_\uparrow - \gamma^5 \psi_\downarrow^c$ ,
which can be used to represent solutions we seek.
Then we proceed with simple algebraic
transformations of the set of four equations (for
$\psi_{\uparrow\downarrow}$, $\psi^c_{\uparrow\downarrow}$,
$\gamma^5 \psi_{\uparrow\downarrow}$ and $\gamma^5
\psi^c_{\uparrow\downarrow}$) to obtain ($\widetilde \nabla_\mu \equiv
\partial_\mu +eB_\mu $)
\begin{eqnarray}
\gamma^\mu \widetilde \nabla_\mu
(\psi_1 - \gamma^5 \psi_4) + ie\gamma^\mu C_\mu (\gamma^5 \psi_2 +\psi_3)
+\kappa (\gamma^5 \psi_2 +\psi_3) &=& 0\, ,\label{eq11}\\ \gamma^\mu
\widetilde \nabla_\mu (\gamma^5\psi_2 + \psi_3) + ie\gamma^\mu C_\mu
(\psi_1 -\gamma^5 \psi_4) +\kappa (\psi_1 -\gamma^5 \psi_4) &=& 0\,
.\label{eq12} \end{eqnarray} Other two equations are obtained after
multiplying (\ref{eq11},\ref{eq12}) by the $\gamma^5$ matrix.\footnote{We
note the interesting fact (obvious and well-known but {\it not always}
fully appreciated):  after different constraints one imposes on the
functions $\psi_{1,2,3,4}$ one obtains different physical pictures.
Namely, if $\psi_2 +\gamma^5 \psi_3 = \psi_4 - \gamma^5 \psi_1$\, ,
which results in the rather unexpected constraints
$\gamma^5 (\psi_\uparrow^c +\psi_\downarrow^c) = 0$,
one can recover the equation which is similar to  the Dirac
equation, the
interaction of the eigenstates of the charge operator with the 4-vector
potential, but with the opposite sign in the mass term.  The complete
consideration of this condition should be presented in a separate paper.
At this point we note only that it appears to express the fact that for
describing a pair of charge particles it is sufficient to use only the
Dirac field operator $\psi (x^\mu) = \psi_\uparrow +\psi_\downarrow$.}
Let us next impose $\widetilde \Psi^S =0$ and $\widetilde \Psi^A =0$.
They are equivalent either to the
constraints on the creation/annihilation operators $a_\uparrow (p^\mu) =
b_\downarrow (p^\mu)$ and $a_\downarrow (p^\mu) = b_\uparrow (p^\mu)$, or
constraints $\psi_\uparrow = \psi_\uparrow^c \equiv \psi^s$ and
$\psi_\downarrow = -\psi_\downarrow^c \equiv \psi^a$. The functions
$\psi_{1,2,3,4}$ become to be interrelated by the conditions
\begin{equation}
\psi_1 = \psi^s -\gamma^5 \psi^a\, ,\,
\psi_2 = \psi^a +\gamma^5 \psi^s\, , \,
\psi_3 \equiv \gamma^5 \psi_1\, , \,
\psi_4 \equiv \gamma^5 \psi_2\, .
\end{equation}
It is the simple procedure to show that $\psi_1$ presents itself
self charge conjugate field and $\psi_2$, the anti-self charge
conjugate field.\footnote{The operator of the charge conjugation
and the chirality $\gamma^5$ operator are {\it anti-commuting}
operators.}
As a result one obtains the following dynamical equations:
\begin{equation}
\gamma^\mu D_\mu^\ast \psi_1 +\kappa \gamma^5 \psi_2 = 0\, ,\quad
\gamma^\mu D_\mu \psi_2 -\kappa \gamma^5 \psi_1 = 0\, ,\label{sa2}
\end{equation}
where the lengthening derivative is now defined
$$D_\mu = \partial_\mu -ie\gamma^5 C_\mu +eB_\mu\, .$$
The equations for the Dirac conjugated counterparts of $\psi_{1,2}$
read
\begin{equation}
\partial_\mu \overline \psi_1 \gamma^\mu +\kappa \overline\psi_2 \gamma^5
= 0\, ,\quad
\partial_\mu \overline \psi_2 \gamma^\mu -\kappa \overline\psi_1 \gamma^5
= 0\, .
\end{equation}
One can propose the Lagrangian for free fields $\psi_{1,2}$
and their Dirac conjugates (cf. with the concept of the extra Dirac
equations in ref.~\cite{DVO0} and  with the spin-1 case,
ref.~\cite{DVOEG}):\footnote{We still leave the room for other
kinds of the Lagrangians describing self/anti-self charge conjugate
states, see below and cf.~\cite{DVO}.}
\begin{eqnarray}
{\cal L}^{free} &=& {1\over 2} \left [ \overline \psi_1 \gamma^\mu
\partial_\mu \psi_1 -\partial_\mu \overline\psi_1 \gamma^\mu \psi_1
+\overline\psi_2 \gamma^\mu \partial_\mu \psi_2 -
\partial_\mu \overline \psi_2 \gamma^\mu \psi_2 \right ]
+ \nonumber\\
&+& \kappa \left [\overline \psi_1 \gamma^5 \psi_2 -\overline \psi_2
\gamma^5 \psi_1 \right ]\, ;
\end{eqnarray}
and the terms of the interaction:
\begin{equation}
{\cal L}^{int} =   ie (\overline \psi_1 \gamma^\mu \gamma^5 \psi_1
-\overline \psi_2 \gamma^\mu\gamma^5 \psi_2 ) C_\mu
+ e (\overline \psi_1 \gamma^\mu \psi_1 +\overline \psi_2 \gamma^\mu
\psi_2 ) B_\mu\, .
\end{equation}
The conclusion that self/anti-self charge
conjugate can possess the axial charge is {\it in
accordance} with the conclusions of
refs.~\cite{ZIINO1,ZIINO2,DVO,ZIINO3} and the old ideas of R. E.
Marshak et al.~\cite{MARSHAK}. It is the remarkable feature of this
model that we did {\it not} assume that self/anti-self charge conjugate
fields are massless.

Now it is natural to ask the question, what physical
excitations and which interaction scheme would we obtain if we impose
different constraints on the positive- and negative- energy solutions.
The Ahluwalia reformulation of the Majorana-McLennan-Case construct was
presented recently~\cite{DVA}. The following type-II spinors have been
defined
\begin{equation}
\lambda^{^{S,A}} = \pmatrix{\zeta_\lambda
\Theta_{[j]} \phi_{_L}^\ast (p^\mu)\cr \phi_{_L} (p^\mu) \cr}\,\, , \quad
\rho^{^{S,A}} = \pmatrix{\phi_{_R}
(p^\mu)\cr (\zeta_\rho \Theta_{[j]} )^\ast \phi_{_R}^\ast (p^\mu)
\cr}\,\, .  \label{type2}
\end{equation}
The positive-energy solutions are presented, e.~g., by the self charge
conjugate $\lambda^{{^S}}$ spinors, the negative energy solutions, by
the anti-self charge conjugate $\lambda^A$ spinors, see the formula (46)
in~\cite{DVA}. In our
choice of the operator of the charge conjugation ($\vartheta_c = 0$)  the
phase factors $\zeta_{\lambda , \rho}$ are fixed as $\pm 1$, for
$\lambda^{{^S}}$ (and $\rho^{{^S}}$), and $\mp 1$, for $\lambda^{{^A}}$
(and $\rho^{{^A}}$), respectively. One can find relations between the
type-II spinors and
the Dirac spinors. They are listed here
\begin{eqnarray}
\lambda^{^{S,A}}_\uparrow (p^\mu) &=& +{1-\gamma^5 \over 2} u_\uparrow
(p^\mu) \pm {1+\gamma^5 \over 2} u_\downarrow (p^\mu)\, ,\\
\lambda^{^{S,A}}_\downarrow (p^\mu) &=& \mp {1+\gamma^5 \over 2} u_\uparrow
(p^\mu) +{1-\gamma^5 \over 2} u_\downarrow (p^\mu)\, .
\end{eqnarray}
By  using these relations one can deduce how is the $\nu$ operator, which
was given by D. V. Ahluwalia, connected with the Dirac field operator and
its charge conjugate. If $(a_\downarrow + b_\uparrow)/2 = (a_\uparrow -
b_\downarrow)/2 \equiv c_\uparrow = d_\downarrow $ and $(a_\uparrow +
b_\downarrow)/2 = (a_\downarrow -b_\uparrow)/2 \equiv c_\downarrow =
d_\uparrow$ one has
\begin{equation} \nu (x^\mu) = +{1\over 2}
(\psi_\downarrow (x^\mu) - \psi^c_\uparrow (x^\mu) ) - {\gamma^5 \over 2}
(\psi_\uparrow (x^\mu) +\psi^c_\downarrow (x^\mu) )\, .  \end{equation}
The operator composed of $\rho$ spinors
\begin{equation}
\tilde \rho (x^\mu) \equiv \int \frac{d^3 \vec p}{(2\pi)^3} {1\over 2p_0}
\sum_\eta \left [ \rho^{{^A}}_\eta (p^\mu) e_\eta (p^\mu) e^{-ip\cdot
x} +
\rho^{{^S}}_\eta (p^\mu) f_\eta^\dagger (p^\mu) e^{+ip\cdot  x}\right ]
\end{equation}
is then expressed\,\footnote{Of course, the certain relations between
creation/annihilation operators of different field operators are again
assumed.}
$\tilde \nu (x^\mu) = +{1\over 2}
(\psi_\downarrow (x^\mu) + \psi_\uparrow^c (x^\mu)) +{\gamma^5 \over 2}
(\psi_\uparrow (x^\mu) -\psi_\downarrow^c (x^\mu))$\, .
Other fields which we use in order to obtain dynamical equations
are $\nu^c (x^\mu)$, $\gamma^5 \nu (x^\mu)$ and $\gamma^5 \nu^c (x^\mu)$,
$\tilde \nu^c (x^\mu)$, $\gamma^5 \tilde \nu (x^\mu)$ and $\gamma^5 \tilde
\nu^c (x^\mu)$. Their explicit forms will be presented in the extended
version of this paper elsewhere.
After rather tiresome calculation procedure one obtains the dynamical
equations in this approach
\begin{eqnarray}
&&\gamma^\mu \widetilde \nabla_\mu (\nu - \nu^c ) +ie\gamma^\mu \gamma^5
C_\mu (\nu - \nu^c ) +\kappa\gamma^5 (\tilde \nu +\tilde\nu^c ) = 0\, ,\\
&&\gamma^\mu \widetilde \nabla_\mu (\tilde \nu + \tilde \nu^c )
-ie\gamma^\mu \gamma^5 C_\mu (\tilde \nu + \tilde \nu^c ) -\kappa\gamma^5
(\nu - \nu^c ) = 0\, ,\\
&&\gamma^\mu \gamma^5 \widetilde \nabla_\mu (\nu - \nu^c )
+ie\gamma^\mu C_\mu (\nu - \nu^c ) -\kappa
(\tilde \nu +\tilde \nu^c ) = 0\, ,\\
&&\gamma^\mu \gamma^5 \widetilde \nabla_\mu (\tilde \nu +\tilde \nu^c )
-ie\gamma^\mu C_\mu (\tilde \nu + \tilde \nu^c ) +\kappa
(\nu - \nu^c ) = 0\, ,
\end{eqnarray}
and, for $\nu_1 = \nu -\nu^c+\tilde\nu+\tilde\nu^c$\, ,\,
$\nu_2 = \nu-\nu^c-\tilde\nu -\tilde\nu^c$,
\begin{eqnarray}
&&(1\pm\gamma^5) \left [ \gamma^\mu \widetilde \nabla_\mu \nu_1
\mp ie\gamma^\mu C_\mu \nu_2 \mp
\kappa \nu_2 \right ] = 0\, ,\\
&&(1\pm \gamma^5) \left [ \gamma^\mu \widetilde
\nabla_\mu \nu_2 \mp ie\gamma^\mu C_\mu
\nu_1 \pm \kappa \nu_1 \right ] = 0\, .
\end{eqnarray}
Thus, one can see that the operators $\nu (x^\mu) -
\nu^c (x^\mu)\, ,\,\,\mbox{and}\,\, \tilde\nu (x^\mu) +\tilde\nu^c
(x^\mu)\,\, $ also satisfy the equations of the type (\ref{sa2}).

The Ahluwalia construct is, in fact, based on other postulates,
which may be used to derive the Dirac equations.\footnote{They are {\it
not} less general than the Dirac postulates, but they can be applied to
any spin constructions. Here they are
\begin{itemize}
\item
The Wigner rules for transformations of left $(0,j)$ and right $(j,0)$
``spinors" of any spin:
\begin{equation}
\phi_{_{R,L}} (p^\mu) = \Lambda_{_{R,L}} (p^\mu \leftarrow
\overcirc{p}^\mu) \phi_{_{R,L}} (\overcirc{p}^\mu )= \exp (\pm \vec J\cdot
\vec \varphi ) \phi_{_{R,L}} (\overcirc{p}^\mu)\, ,
\end{equation}
with $\vec J$ being the spin
matrices for spin $j$, and $\varphi$ being the parameters of the boost.
The $(0,j)$- spinors (and $(j,0)$- spinors) are assumed to be in the
helicity eigenstates.

\item
The relations between left- and right- ``spinors" in the frame with zero
momentum. For the spin-1/2 states they read
\begin{equation}
\left [\phi_{{_L}}^h (\overcirc{p}^\mu) \right ]^\ast =
(-1)^{1/2 - h} \exp (-i (\theta_1 + \theta_2) ) \Theta_{[1/2]}
\phi_{{_L}}^{-h} (\overcirc{p}^\mu)\, ,
\end{equation}
$\overcirc{p}^\mu = (E=m, \vec 0)$ denotes the zero-momentum
frame; and $\Theta_{[1/2]} = -i\sigma_2$ is the Wigner operator for
spin $j=1/2$.  The property $\Theta_{[j]} \vec J \Theta_{[j]}^{-1} = -
\vec J^{\,\ast}$ is used in derivations of dynamical equations.

\end{itemize}
}
On using his postulates we need not to assume any constraints on
the creation/annihilation operators for self/anti-self charge conjugate
states $c_\eta (p^\mu)$, $d_\eta (p^\mu)$, $e_\eta (p^\mu)$ and $f_\eta
(p^\mu)$. The self charge conjugate $\lambda^{{^S}}$ (anti-self
charge conjugate $\lambda^{{^A}}$) is coupled with anti-self charge
conjugate $\rho^{{^A}}$ (self charge conjugate $\rho^{{^S}}$),
respectively, through the mass term.   Equations become the
eight-component equations both for states (and corresponding field
operators) describing positive-energy    and for the states corresponding
to the negative energy.  See equations (17,18) in~\cite{DVO}.

In the Majorana-like constructs the change of the
phase as $e^{i\alpha}$ would destroy the self/anti-self charge conjugacy.
But, it is seen from (\ref{type2}) that the transformations which  include
the $\gamma^5$ matrix are indeed permitted for the type-II spinors.
The $\gamma^5$ gauge-invariant
Lagrangian was proposed in ref.~\cite{DVO} and local gradient
transformations have been discussed there and in subsequent papers.
Surprisingly, from this formulation one can still
recover~\cite{DVOAPP} the Feynman-Gell-Mann scheme~\cite{FG} for
two-spinors. Furthermore, in the recent paper~\cite{DVOAPP} we
proposed the $SU(2)$ graded transformations for neutrino (self/anti-self
charge conjugate states).  The Majorana-like $\nu$ operator admits the
transformations $\nu^\prime (x^\mu) = (c_0 + i\, \vec \tau\cdot \vec c)
\nu^\dagger (x^\mu)$\, , with $\vec \tau$ generators answering
for the concept of the Wigner {\it sign} spin. If one considers the $q$-
number theory the Lagrangian  remains invariant under this sort of
transformations.

Finally, the question of the existence of ``longitudinal" interactions.
The existence of this type of interactions has been claimed
in the works of M. Evans. Unfortunately, his work contains
many algebraic errors in derivation procedures, so it is difficult for me
to comment it.  But, recently, S. Esposito and E.
Recami~\cite{Espos} found mathematically rigorously the interaction
term of a Dirac fermion with $[\vec A \times \vec A^\ast ]$. In their
formulation which is based only on the usual Dirac equation this term
can be reduced to the term of the type $(\vec \sigma \cdot \vec \nabla )
{\cal V}$, where ${\cal V}$ is a scalar potential. So, it is still of the
interest to investigate the nature of these terms. The possibility of
$\vec \sigma \cdot [\vec A \times \vec A^\ast ]$ terms appears to be also
related with the matters of chiral interactions.  If the Dirac theory
would admit the chiral phase transformations we would be able to write
\begin{equation}
\left [ \gamma^\mu (\partial_\mu -ieC_\mu +e\gamma^5
B_\mu) +\kappa \right ] \psi_1 =0\quad.
\end{equation}
After re-writing
the equation into the two-component form we obtain
\begin{eqnarray}
\sigma^\mu [ \partial_\mu -ie (C_\mu - iB_\mu ) ] \chi +\kappa\phi &=& 0\,
,\label{ci1}\\
\widetilde\sigma^\mu [ \partial_\mu -ie (C_\mu + iB_\mu ) ]
\phi +\kappa\chi &=& 0\, ,\label{ci2}
\end{eqnarray}
with $\psi
=\mbox{column} (\phi\quad \chi)$ and $\sigma$ matrices slightly differing
from the ordinary formulation:  $\sigma^{\mu=0} = \tilde \sigma^{\mu=0}
\rightarrow -i I , \,\, \mbox{and} \,\,\sigma^{\mu=i} = -\tilde
\sigma^{\mu=i} \rightarrow i\sigma^i_{Pauli}$.  From the equation
(\ref{ci1},\ref{ci2}) one can extract the term $\sim e^2 \sigma^i_{Pauli}
\sigma^j_{Pauli} A_i^\ast A_j \, \phi $\, , which leads to the needed
term ($A_\mu = C_\mu +iB_\mu$ again).  But, we are unable to use the
$\gamma^5$ gradient transformation for the Dirac fermion unless consider
another `Dirac-like' equation with the opposite sign in the mass term
$[\gamma^\mu \partial_\mu - \kappa ] \psi_2 =0$\, .  Only in the case of
{\it two} Dirac fields one can construct gauge-invariant Lagrangian. It is
on this formulation that M.  Markov~\cite{MARKOV1,MARKOV2} and
A. Barut and  G.  Ziino~\cite{ZIINO1,ZIINO2,ZIINO3} insisted.  We
still leave the proof of the {\it necessity} of {\it two} Dirac equations
for the extended version of the paper.

%\acknowledgements
{\bf Acknowledgements.} I acknowledge discussions with
Profs. D. V. Ah\-lu\-wa\-lia, A. E. Chu\-by\-ka\-lo, S. Esposito, A. F.
Pashkov and E.  Recami. I am grateful to Zacatecas University, M\'exico,
for a professorship.  This work has been partly supported by the Mexican
Sistema Nacional de Investigadores, the Programa de Apoyo a la Carrera
Docente and by the CONACyT, M\'exico under the research project 0270P-E.

\end{document}